\documentclass[pra, aps, twocolumn, superscriptaddress]{revtex4}

\usepackage[dvips]{graphicx}
\usepackage{color}
\usepackage{amsfonts}
\usepackage{amsmath}

\begin{document}

% MACROS
%%%%%%%%%%%%%%%%%%%%%%%%%%%%%%%%%%%%%%%%%%%%%%%%%%%%%%%%%%%%%%%%%%%%%

\newcommand{\beq}{\begin{equation}}
\newcommand{\eeq}{\end{equation}}
\newcommand{\beqa}{\begin{eqnarray}}
\newcommand{\eeqa}{\end{eqnarray}}
\newcommand{\lf}{\hfil \break \break}
\newcommand{\ahat}{\hat{a}}
\newcommand{\adag}{\hat{a}^{\dagger}}
\newcommand{\adagg}{\hat{a}_g^{\dagger}}
\newcommand{\bhat}{\hat{b}}
\newcommand{\bdag}{\hat{b}^{\dagger}}
\newcommand{\bdagg}{\hat{b}_g^{\dagger}}
\newcommand{\chat}{\hat{c}}
\newcommand{\cdag}{\hat{c}^{\dagger}}
\newcommand{\Pihat}{\hat{\Pi}}
\newcommand{\rhohat}{\hat{\rho}}
\newcommand{\shat}{\hat{\sigma}}
\newcommand{\ket}[1]{\mbox{$|#1\rangle$}}
\newcommand{\bra}[1]{\mbox{$\langle#1|$}}
\newcommand{\ketbra}[2]{\mbox{$|#1\rangle \langle#2|$}}
\newcommand{\braket}[2]{\mbox{$\langle#1|#2\rangle$}}
\newcommand{\bracket}[3]{\mbox{$\langle#1|#2|#3\rangle$}}
\newcommand{\mat}[1]{\overline{\overline{#1}}}
\newcommand{\hak}[1]{\left[ #1 \right]}
\newcommand{\vin}[1]{\langle #1 \rangle}
\newcommand{\abs}[1]{\left| #1 \right|}
\newcommand{\tes}[1]{\left( #1 \right)}
\newcommand{\braces}[1]{\left\{ #1 \right\}}
\newcommand{\sub}[1]{{\mbox{\scriptsize #1}}}
\newcommand{\com}[1]{\textcolor{red}{[\textit{#1}]}}
\newcommand{\novel}[1]{\textcolor{blue}{#1}}

% TITLE
%%%%%%%%%%%%%%%%%%%%%%%%%%%%%%%%%%%%%%%%%%%%%%%%%%%%%%%%%%%%%%%%%%%%%

\title{Amplification of realistic Schr\"{o}dinger cat-like states by homodyne heralding}

\author{Amine Laghaout}
\affiliation{Department of Physics, Technical University of Denmark, Building 309, 2800 Lyngby, Denmark}
\author{Jonas S. Neergaard-Nielsen}
\affiliation{Department of Physics, Technical University of Denmark, Building 309, 2800 Lyngby, Denmark}
\author{Ioannes Rigas}
\affiliation{Quantum Information Theory Group, Institut f{\"u}r Theoretische Physik I, and Max-Planck Research Group, Institute of Optics, Information and Photonics, Universit{\"a}t Erlangen-N{\"u}rnberg, Staudtstra{\ss}e 7/B2, 91058 Erlangen, Germany}
\author{Christian Kragh}
\affiliation{Department of Physics, Technical University of Denmark, Building 309, 2800 Lyngby, Denmark}
\author{Anders Tipsmark}
\affiliation{Department of Physics, Technical University of Denmark, Building 309, 2800 Lyngby, Denmark}
\author{Ulrik L. Andersen}
\affiliation{Department of Physics, Technical University of Denmark, Building 309, 2800 Lyngby, Denmark}
\affiliation{Quantum Information Theory Group, Institut f{\"u}r Theoretische Physik I, and Max-Planck Research Group, Institute of Optics, Information and Photonics, Universit{\"a}t Erlangen-N{\"u}rnberg, Staudtstra{\ss}e 7/B2, 91058 Erlangen, Germany}

\date{\today}

\begin{abstract}
We present a scheme for the amplification of Schr\"{o}dinger cats that collapses two smaller states onto their constructive interference via a homodyne projection. We analyze the performance of the amplification in terms of fidelity and success rate when the input consists of either exact coherent state superpositions or of photon-subtracted squeezed vacua. The impact of imprecise homodyne detection and of impure squeezing is quantified. We also assess the scalability of iterated amplifications.
\end{abstract}

\maketitle

Coherent state superpositions, or optical Schr\"{o}dinger cat states, are widely recognized as promising resources in quantum information \cite{Cochrane1999, Neergaard-Nielsen2010, Ralph2003, Gilchrist2004, Tipsmark2011}, quantum metrology \cite{Ralph2002, Munro2002, Joo2011}, and fundamental tests \cite{Jeong2008, Stobinska2007, Lee2009, McKeown2011}. In the near-orthogonal basis of coherent states $ \braket{\gamma}{{-\gamma}} = e^{-2\gamma^2}$, two particular instances for these states are
\beq
\ket{\kappa_{\pm}(\gamma)} = \frac{1}{\sqrt{2 \pm 2e^{-2\abs{\gamma}^2}}} \tes{\ket{\gamma} \pm \ket{{-\gamma}}},
\label{eq:CatStateDefinition}
\eeq
where the sign ($\pm$) of the superposition refers to the even and odd cat state, respectively. These states exhibit quasi-probability distributions in phase space which are distinctly non-classical. This makes them all the more challenging to generate deterministically as that would require strong Kerr-type non-linearities \cite{Yurke1986, Mecozzi1987, Jeong2004}. One has then to resort to heralding techniques which, though probabilistic, need only linear optics and projective measurements \cite{Glancy2008}. These state-engineering schemes are nonetheless approximative and present a limitation in the fidelity they produce with ideal cat states. Photon-subtraction of squeezed vacuum, for example, is a well-established method to generate approximations of small amplitude cat states, colloquially referred to as Schr\"{o}dinger \textit{kittens} \cite{Dakna1997, Wenger2004, Ourjoumtsev2006, Gerrits2010, Neergaard-Nielsen2006}. Even in the best experimental conditions, the fidelity between the photon-subtracted squeezed vacuum (PSSV) and an actual odd cat state $\ket{\kappa_{-}(\gamma)}$ degrades markedly for $\gamma \geq 1.2$ \cite{Jeong2005}. Yet, for these states to be reliable resources in quantum computation, their fidelity with cat states at least as large as $\gamma = 1.2$ need to be maintained at near-unit fidelity \cite{Ralph2003, Lund2008}. 

Single-photon subtraction is only one example of several measurement-induced schemes which have been proposed to generate kitten states \cite{Ourjoumtsev2007, Wakui2007, Gerrits2010, Nielsen2007, Takeoka2008, Marek2008, Lee2012}. However, none of these schemes can produce arbitrarily large cats in a single run. Ways to get around this issue have been devised using the recursive amplification of small, approximate cats \cite{Lund2004, Takeoka2007}. For example, it was suggested in \cite{Brask2010} to interfere a supply of delocalized single photons followed by homodyne heralding to generate large entangled cat states. These proposals have in common that they rely on the coherent mixing of two small cats, whereupon a projective measurement collapses one of the two outputs onto a \textit{constructive interference} of the inputs---hence the amplification. Here, we shall pursue the same idea but make use solely of homodyne heralding for its relative simplicity and high quantum efficiency. We also demonstrate that the acceptance window of homodyne heralding can be widened to increase the success rate of the amplification while at the same time maintaining a satisfactory fidelity at the output.

The outline of this article is as follows. In Sec. \ref{sec:KittenGeneration} we review the generation of odd kitten states from squeezed vacuum. The output is compared to the ideal odd cat state and the effects of impure squeezing are illustrated. In Sec. \ref{sec:CatAmplification} we present the amplification scheme in the case of ideal input cats and model the effect of a wide homodyning window. Sec. \ref{sec:KittenAmplification} then considers the amplification of the more realistic PSSV. The impact of both impure squeezing and wide post-selection is illustrated. In \ref{sec:Scalability}, we return to the idealized case of ideal homodyning and pure squeezing to consider how our scheme scales with iterated amplifications.

% Approximation of small odd cats
%%%%%%%%%%%%%%%%%%%%%%%%%%%%%%%%%%%%%%%%%%%%%%%%%%%%%%%%%%%%%%%%%%%%%
\section{Approximation of small odd cats}
\label{sec:KittenGeneration}

\begin{figure}[h]
  \includegraphics{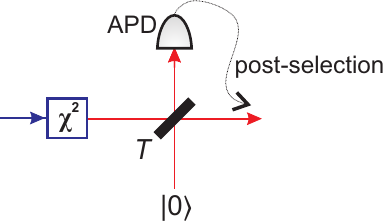}
  \caption{(Color online) Setup for the generation of approximate small odd cat states. A squeezed vacuum, represented here by the pumping of a $\chi^2$  nonlinear medium, is partially reflected onto an ``on-off'' photon detector such as an avalanche photodiode (APD). Upon the reflection of at least one photon from the squeezed vacuum, a state that likens the squeezed single photon is prepared in the limit of unit transmission $T \rightarrow 1$.}
	\label{fig:kitten_setup}
\end{figure}

In this section, we shall briefly review the generation of odd kitten states from the photon-subtraction of squeezed vacuum and analyze its performance in the face of impure squeezing. The basic setup is depicted in Fig. \ref{fig:kitten_setup}. The original proposal of Dakna \textit{et al.} \cite{Dakna1997} required that the photon subtraction be performed by photon-number resolving detectors. However, as is done in most practical schemes \cite{Wenger2004, Neergaard-Nielsen2006, Ourjoumtsev2006}, we shall assume that the post-selection is a binary detection of either zero photons or at least one photon, as would be allowed for by an avalanche photodiode (APD). The modeling of such an ``on-off'' post-selection operation is presented in Sec. \ref{sec:GaussianCircuitry}.

Quantum inefficiencies and dark counts are two nuisances inherent to photodetection which should be reckoned with. Whereas the former merely affects the success probability of the scheme, the latter contaminates the output with a squeezed vacuum component which weakens the nonclassicality of the output. An equally detrimental effect is the impurity of the squeezed vacuum. We shall not treat quantum inefficiencies and dark counts here as they have already been covered in \cite{Glancy2008} in the context of cat amplification. We will however look at the fidelity response to the amplitude and impurity of squeezing. 

% Fidelity of the output for pure squeezing
%%%%%%%%%%%%%%%%%%%%%%%%%%%%%%%%%%
\subsection{Fidelity of the output for pure squeezing}

\begin{figure}[ht]
	\begin{tabular}{ c c c c c }
  	\includegraphics[width=.8\columnwidth]{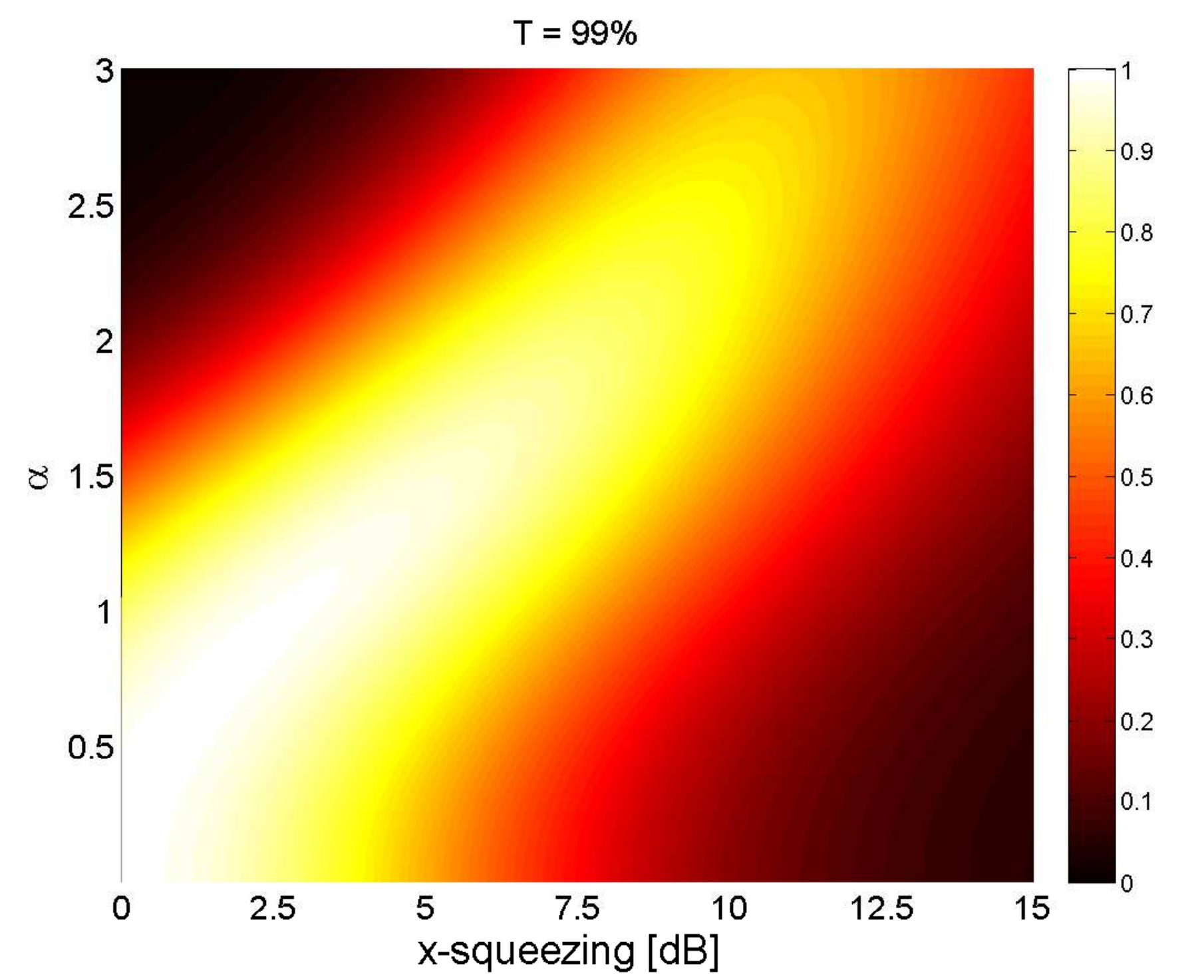} \\\\
	\includegraphics[width=.9\columnwidth]{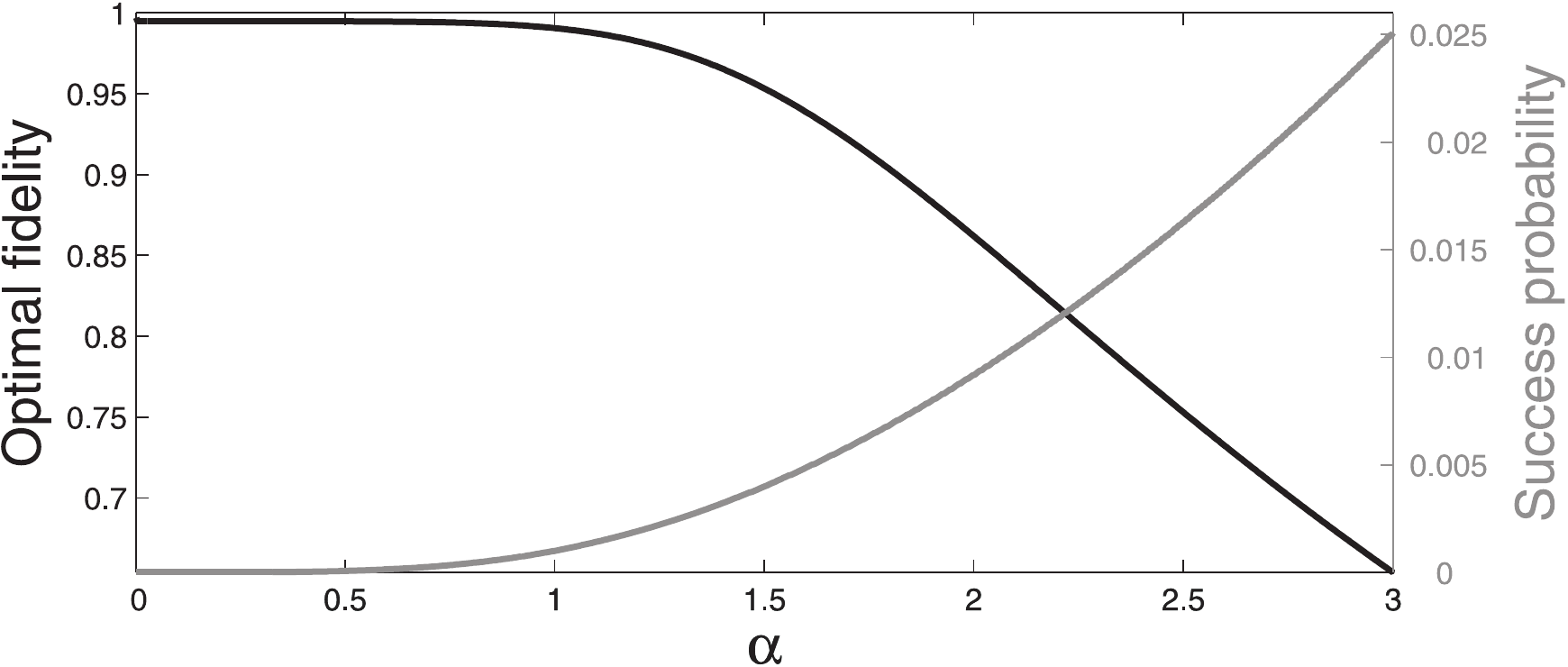}
  \end{tabular}
  \caption{(Color online) \textit{Top}: Contour plot of the fidelity of the PSSV state with $\ket{\kappa_{-}(\alpha)}$ given an input squeezing $\xi_x$ between 0 and 15 dB. \textit{Bottom}: Maximum achievable fidelity of the PSSV with $\ket{\kappa_{-}(\alpha)}$ (black) and corresponding success probability as a function of $\alpha$ (gray). In both cases, the transmission of the subtraction beam splitter is 99\%.}
  \label{fig:KittenFidelity}
\end{figure}

Let's assume from now on that the squeezed vacuum is pure and anti-squeezed in the \textit{x}-quadrature. I.e., if we denote the squeezing relative to the shot noise variance by $\xi$ (in dB), then $\xi_x = -\xi_p > 0$, where the subscript labels the measured quadrature \footnote{For simplicity, we shall refer to both squeezing and anti-squeezing as \textit{squeezing} $\xi$. The difference between the two is only made by the sign of $\xi$}. The fidelity of the PSSV state with an ideal cat state $\ket{\kappa_{-}(\alpha)}$ then depends on the squeezing $\xi$ and the transmission $T$ of the subtraction beam splitter \footnote{The transmission of the subtraction beam splitter does not just affect the success probability of the scheme, but also the fidelity of the output state. This is because we are using an ``on-off'' post-selection where we assume that by setting $T \rightarrow 1$, only one photon will makes its way to the APD. A lower transmission would furthermore inflict higher loss of the initial squeezed state.}. The effect of these parameters is illustrated in Fig. \ref{fig:KittenFidelity} where we can see that the fidelity is optimized for low squeezing and near-unit beam splitter transmission $T \rightarrow 1$. For any given input squeezing, there corresponds a finite amplitude $\alpha$ of the target cat with which the output has an maximized fidelity. For example, a squeezing of, say, 3 dB is optimal for producing an approximation to a cat state of size $\alpha = 1.0$. A complementary investigation of PSSV states that looked at nonclassicality as the main figure of merit (as opposed to fidelity) is given by Kim \textit{et al.} in \cite{Kim2005}.

In order to assess the performance of the PSSV generation, it is most instructive to restrict ourselves to the maximal fidelity achievable with any target cat of size $\alpha$. This fidelity optimum is traced by the ridge of Fig. \ref{fig:KittenFidelity} (top) and its $\alpha$ dependence is reproduced in Fig. \ref{fig:KittenFidelity} (bottom) along with the corresponding success rate. For example, if one wants to produce an approximate odd cat state of size $\alpha = 1.5$, the required squeezing should be $\xi \approx 5.2$ dB for a fidelity of at most 95.4\% and a success probability of 0.4\%.

As far as the success probability is concerned, it can be increased at the expense of fidelity by increasing the incidence of photons on the APD via stronger squeezing or weaker beam splitter transmission.

% Impact of impure squeezing
%%%%%%%%%%%%%%%%%%%%%%%%%%%%%%%%%%
\subsection{Impact of impure squeezing} 

In any real world experiment, noise will inevitably inflate the variance of the squeezed vacuum in either quadrature. This impurity has been explained as stemming from losses or from multimode parametric down-conversion \cite{Tualle-Brouri2009} whereby the photons triggering the post-selection belong to a different spatial or frequency mode than the heralded state. Impure squeezing may originate in the down-converter itself or, more generally, at any point in the setup where vacuum contamination or modal mismatch could take place, including at the detectors  (e.g., via quantum inefficiencies). Regardless of its root causes, we shall wrap these impurities into a single parameter $\epsilon$ relating the squeezing $\xi$ in dB of the \textit{x}- and \textit{p}-quadratures,
\beq
\xi_x = -\epsilon \xi_p,
\label{eq:epsilonDefinition}
\eeq
whereby pure squeezing corresponds to $\epsilon = 1$. As derived in \S \ref{sec:PuritySqueezedVacuum}, the purity of the squeezed vacuum is given by 
\beq
\mathcal{P} = 10^{-\frac{1}{20} \tes{1-\epsilon} \xi_p}.
\eeq
Note that the Heisenberg uncertainty relation imposes that $\epsilon \geq 1$.

\begin{figure}[h]
  \includegraphics[width=.8\columnwidth]{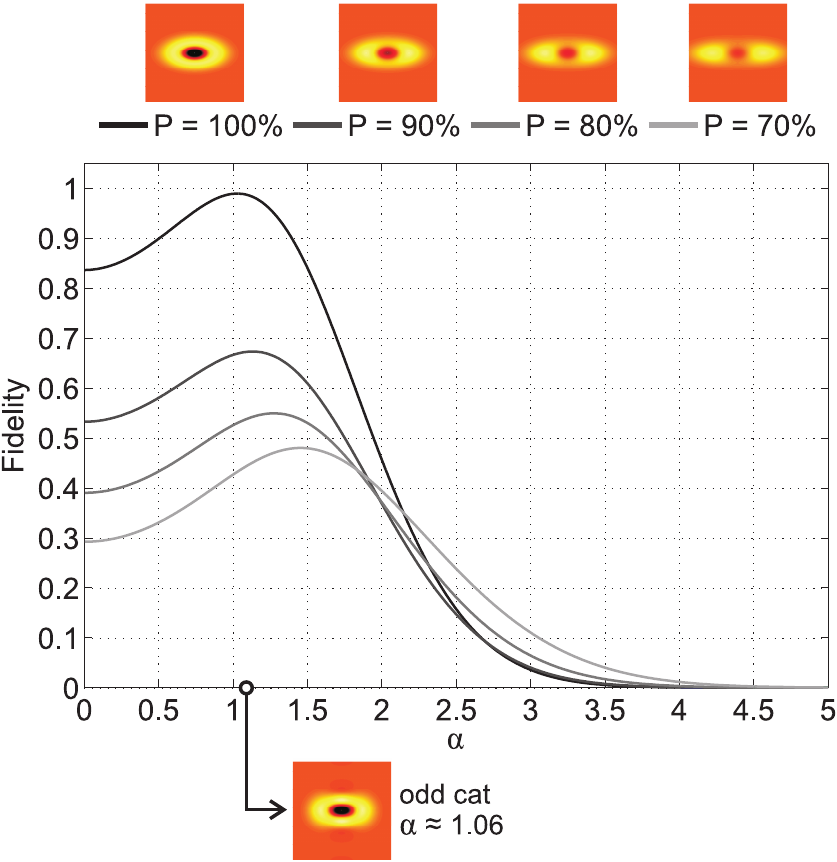}
  \caption{(Color online) Fidelity between an ideal cat state $\ket{\kappa_{-}(\alpha)}$ and the kitten states obtained from squeezed vacuum $\xi_p = -3$ dB at four squeezing purities $\mathcal{P}$ (marked in decreasing darkness of gray): 100\%, 90\%, 80\% , and 70\%. The Wigner functions of the PSSVs corresponding to these four purities are shown at the top. As a reference of what the ``ideal'' output ought to look like, the Wigner function of the state $\ket{\kappa_{-}(1.06)}$, which has the highest fidelity with the PSSV obtained from pure squeezing, is shown at the bottom. The transmission of the subtraction beams splitter is set at 99\%.}
\label{fig:ImpureFidelity}
\end{figure}

The impact of impurity on fidelity is plotted in Fig. \ref{fig:ImpureFidelity} where we set the squeezing at $-3.0$ dB and adjust the anti-squeezing according to four different settings of purity. The immediate observation is that a decrease of 10\% in purity, from 100\% to 90\%---corresponding to an increase of anti-squeezing to $+3.9$ dB from $+3.0$ dB, leads to a drop of the maximum fidelity of nearly 32\%.

% Amplification of ideal cats
%%%%%%%%%%%%%%%%%%%%%%%%%%%%%%%%%%%%%%%%%%%%%%%%%%%%%%%%%%%%%%%%%%%%%
\section{Amplification of ideal odd cats}
\label{sec:CatAmplification}

\begin{figure}[h]
  \includegraphics[width=1\columnwidth]{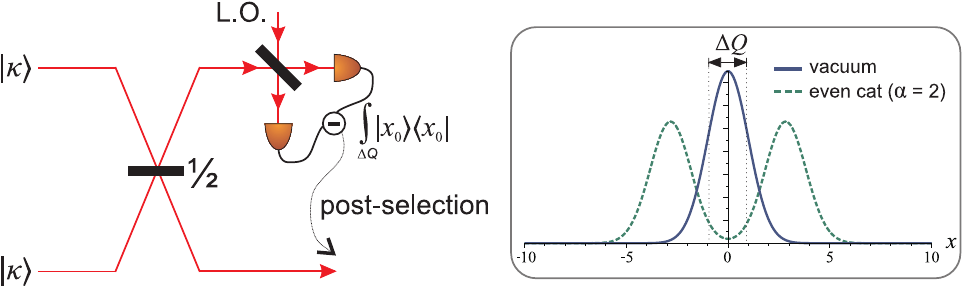}
  \caption{(Color online) Setup for the amplification of two ideal Schr\"{o}dinger cats into a larger even cat. The two inputs are mixed on a symmetric beam splitter and  one of the outputs is projected onto an \textit{x}-quadrature window of width $\Delta Q$ around $x_0 = 0$ by an otherwise ideal homodyne detector. \textit{Inset}: Wave functions $\braket{x}{0}$ (solid curve) and $\braket{x}{\kappa_{+}(2)}$ (dashed curve) of the vacuum and of an even cat state of size $\alpha = 2$, respectively. The two functions are best distinguished at $x = 0$ where their overlap is minimized.}
\label{fig:amplification_setup}
\end{figure}

In Fig. \ref{fig:amplification_setup} we present our amplification setup: Two identical cats are mixed on a balanced beam splitter whereupon one of the ensuing modes heralds the amplified output based on the measurement of \textit{x}-quadratures around \textit{x} = 0. The scenario where the inputs are idealized cats of opposite parity has  been outlined by Takeoka and Sasaki in \cite{Takeoka2007}. We will however look at the more practical case where the input have identical parity. 

Let's briefly run through the evolution of the state in setup. We can readily see that the state emerging from the first balanced beam splitter contains an even cat of amplitude $\sqrt{2}$ times larger:
\beqa
& & \ket{\kappa_{\pm}(\alpha)} \otimes \ket{\kappa_{\pm}(\alpha)} \nonumber\\
& \rightarrow &  \ket{\kappa_{+}(\sqrt{2}\alpha)} \otimes \ket{0} \pm \ket{0} \otimes \ket{\kappa_{+}(\sqrt{2}\alpha)}.
\label{eq:PsiOutAfterBS}
\eeqa

If, by measuring one of the modes, we can post-select $\ket{0}$ from $\ket{\kappa_{+}(\sqrt{2}\alpha)}$, then the other mode will collapse onto the desired amplified state $\ket{\kappa_{+}(\sqrt{2}\alpha)}$. The accuracy of this discrimination is of course limited by the overlap of vacuum with the cat $\braket{0}{\kappa_{+}(\sqrt{2}\alpha)}$, which is however negligible for $\alpha \gg 0$. Based on the wave function profiles of the vacuum and the cat, we can see that homodyne measurement of the $x = 0$ quadrature is indeed a good way to discriminate the two states as it is where their overlap is minimized (cf. inset of Fig. \ref{fig:amplification_setup}). The ambiguity of the discrimination is quantified in \S \ref{sec:AmbiguousProj}.

Applying the homodyne projection $\ketbra{x_2 = 0}{x_2 = 0}$ on the second mode of (\ref{eq:PsiOutAfterBS}), we are left with 
\beq
\ket{\psi_\sub{out}} = \frac{e^{\alpha^2}}{2} \cdot  \frac{\ket{\sqrt{2}\alpha} + \ket{{-\sqrt{2}\alpha}} \pm 2 e^{-2\alpha^2} \ket{0}}{\sqrt{\cosh(2\alpha^2) + e^{-2\alpha^2} \pm 2e^{-\alpha^2}}}.
\label{eq:AmplifiedCat}
\eeq
The fidelity of this state with an ideal even cat state of size $\beta$ is
\beq
F = \mbox{sech}(\beta^2) \cdot \frac{\abs{\cosh(\sqrt{2}\alpha\beta) \pm e^{-\alpha^2}}^2}{\cosh(2\alpha^2) + e^{-2\alpha^2} \pm 2e^{-\alpha^2}}.
\eeq

In practice, a valid output is heralded whenever the homodyne detector records a state whose  \textit{x}-quadrature lies within a window $\Delta Q$ around $x = 0$ where $\Delta Q$ shall be expressed in shot noise units (SNUs) \footnote{By shot noise unit, we mean the standard deviation of shot noise in phase space. By setting $\hbar = 1$, this implies $SNU = \frac{1}{\sqrt{2}}$ (in absolute units of phase space quadrature).}. We keep track of such a post-selection window because no practical homodyning device has enough resolution to truly project onto an exact quadrature $\ketbra{q_0}{q_0}$. Experimentally, such a precise projection would not be desirable either for it would lead to very small success probabilities. A compromise is therefore to allow a finite post-selection range. A full model of this realistic ``wide'' homodyning is presented in Sec. \ref{sec:WideHomodyning}.

\begin{figure}[!ht]
  \includegraphics[width=.8\columnwidth]{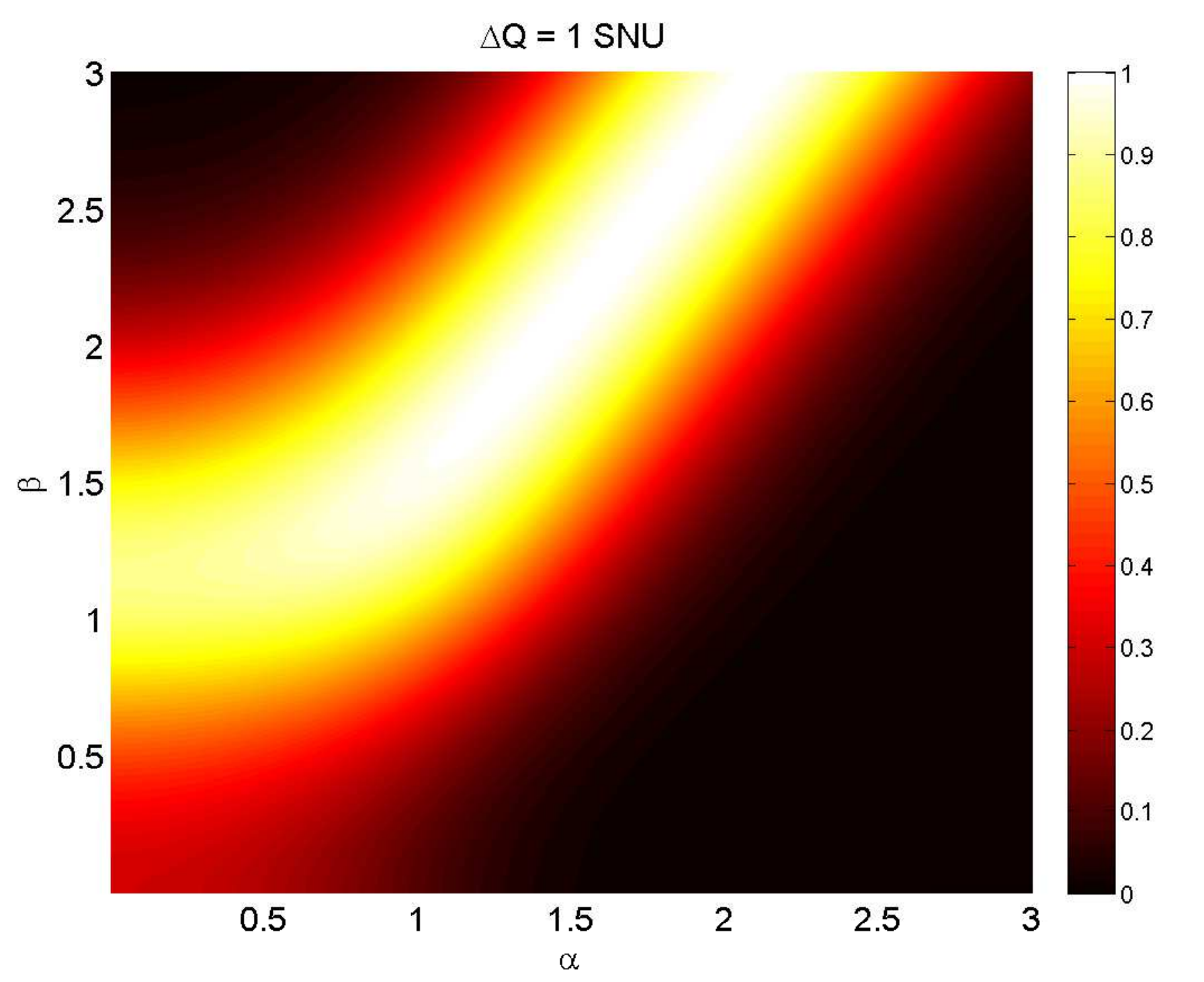}
  \caption{(Color online) Contour plot of the fidelity between an even cat of size $\beta$ and the amplification obtained from two odd cats of size $\alpha$. The post-selection window is set to 1 SNU.}
\label{fig:CatAmplification}
\end{figure}

From now on, we shall only consider odd cat inputs to the amplification setup. (The next section deals with approximations to odd cat inputs, namely PSSVs.) A contour plot of the fidelity between an even cat of size $\beta$ and the amplified state from two odd cats of size $\alpha$ is shown in Fig. \ref{fig:CatAmplification} for a homodyning window of $1$ SNU. The $\sqrt{2}$ amplification factor that was also witnessed in earlier schemes \cite{Lund2004, Suzuki2006, Takeoka2007} is recognizable as the slope $\frac{\beta}{\alpha}$ where the fidelity is optimized. The bend of this optimum crest for $\alpha \leq 1$ arises from the vacuum component which ``survives'' the post-selection but vanishes from the output state $\ket{\psi_\sub{out}}$ for larger $\alpha$. This feature is in a sense a manifestation of the discreteness of photon numbers for weak coherent states. As can be seen from (\ref{eq:AmplifiedCat}), the output consists of an even cat minus a vacuum. For $\alpha \rightarrow 0$, this ``subtraction'' of the vacuum component yields a state whose 2-photon component has a relatively higher weight than in any even cat of size $\beta \ll 1$. The proportionality in amplitude between input and amplified cats thus breaks down, and it is instead $\ket{\beta \approx 1}$---of all even cats---that exhibits the best fidelity with the output.

\begin{figure}[h]
  \includegraphics[width=1\columnwidth]{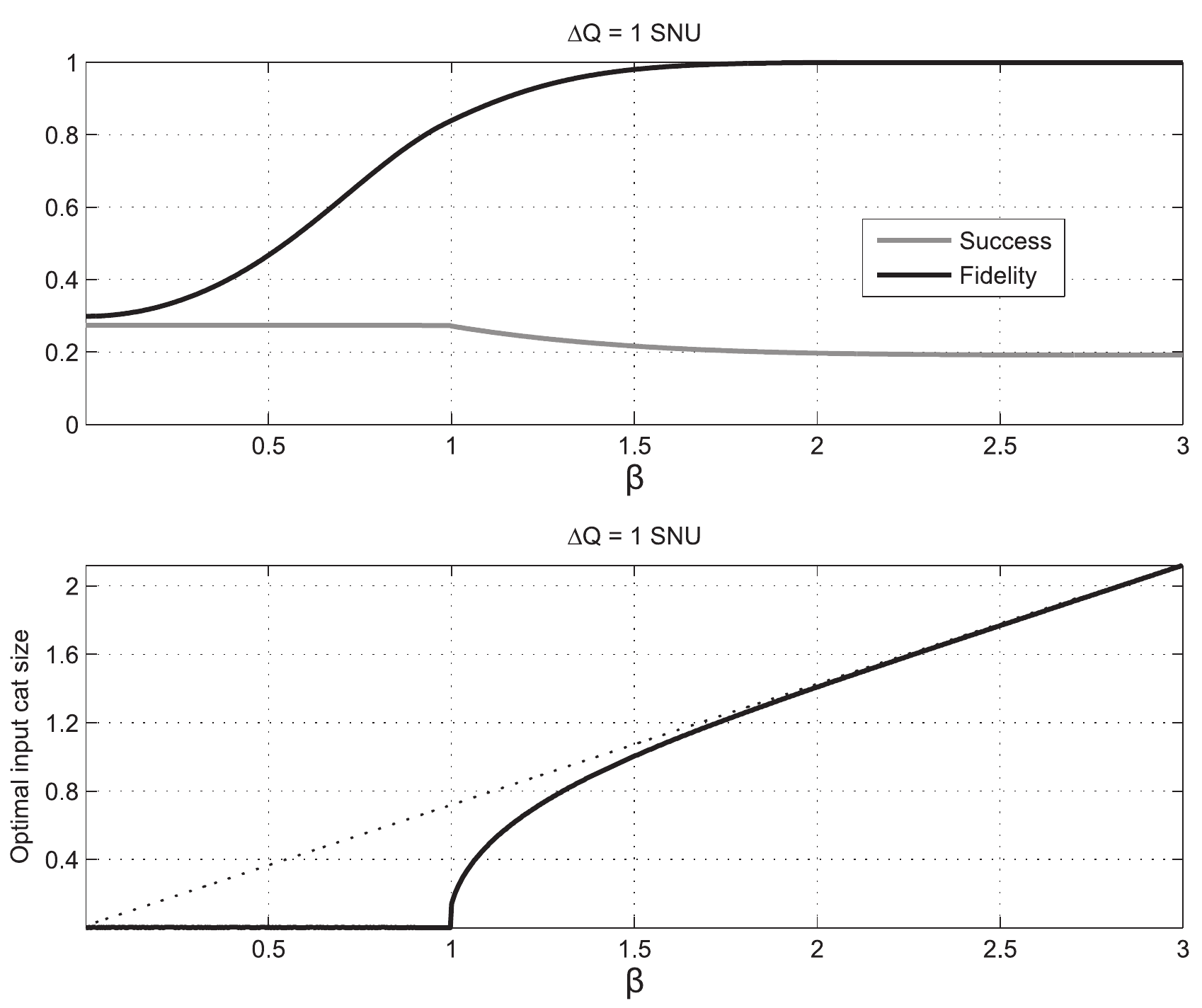}
  \caption{\textit{Top}: Maximum fidelity (black) and corresponding success probability (gray) of the amplified output with respect to an even cat state of size $\beta$. \textit{Bottom}: Amplitude $\alpha$  of the input required to obtain the maximum fidelity of the output with an even cat state of size $\beta$. The dotted line marks the $\sqrt{2}$ amplification factor. The homodyne post-selection window is 1 SNU wide.}
\label{fig:CatAmplificationOptimal}
\end{figure}

Just as in Fig. \ref{fig:KittenFidelity} for the PSSV state, it is informative to look at the mapping between input and output parameters which optimizes fidelity. This is shown in Fig.  \ref{fig:CatAmplificationOptimal} where the optimal input state amplitude is plotted as a function of the amplitude $\beta$ of the target state $\ket{\kappa_{+}(\beta)}$. For example, if one wants to produce an even cat state of size $\beta = 2$, then an input odd cat state of size $\alpha = 1.4$ is required. The fidelity of the output with $\ket{\kappa_{+}(\beta = 2)}$ will be nearly $100\%$ and the success probability of the operation about $20\%$. As discussed above, the flat plateau for $\beta \leq 1$ corresponds to the range where the vacuum component that filtered through the post-selection
becomes predominant. The consequence is that the single-photon ``cat state'', $\ket{\kappa_{-}(\alpha \to 0)} \approx \ket{1}$, becomes the only input to optimize outputs of target size $\beta \in [0, 1]$.

\begin{figure}[h]
  \includegraphics[width=1\columnwidth]{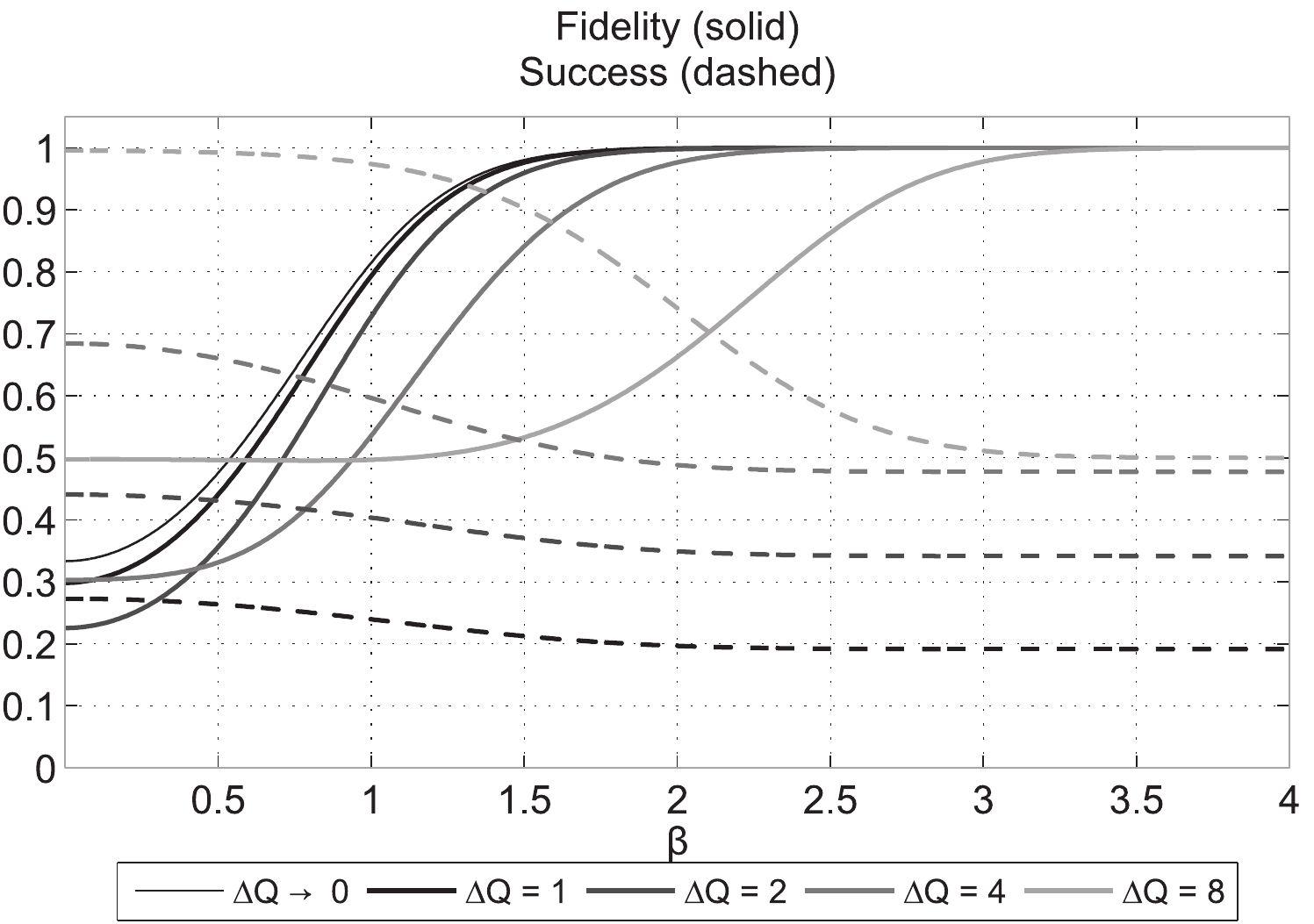}
  \caption{\textit{Solid curves}: Fidelity between an even cat of size $\beta = \sqrt{2}\alpha$ and the output of mixing two odd cats of size $\alpha$ for five homodyning widths $\Delta Q$ (in SNUs). \textit{Dashed curves}: Corresponding success probabilities. (The curve for $\Delta Q \to 0$ is zero throughout since the probability of picking out the exact $x = 0$ quadrature is vanishingly small.) The shading of the curves is lightens with larger homodyning windows.}
\label{fig:HDwindowRobustness}
\end{figure}

To assess the robustness of the amplification scheme to the homodyning width $\Delta Q$, Fig. \ref{fig:HDwindowRobustness} plots the fidelity between the output produced from two odd cats $\ket{\kappa_{-}(\alpha)}$ and an even cat $\ket{\kappa_{+}(\beta = \sqrt{2}\alpha)}$ given different homodyning windows. The fidelity curve for exact homodyning $\Delta Q \to 0$ is also plotted as a reference. It can be seen that the amplification is only vulnerable to $\Delta Q$ for small inputs. Beyond a target size of $\beta \approx 3.5$, homodyning widths of up to 8 SNUs hardly have any effect on the fidelity. From an experimental point of view, this robustness of the homodyning post-selection allows one to reach higher success probabilities without compromising fidelity. For example, by sending in two odd cats of size $\alpha = 2.5$, one every two homodyne measurements will successfully herald an even cat of size $\beta = \sqrt{2} \times 2.5 \approx 3.5$ with a fidelity of nearly 100\%.

\begin{figure}[h]
  \includegraphics[width=1\columnwidth]{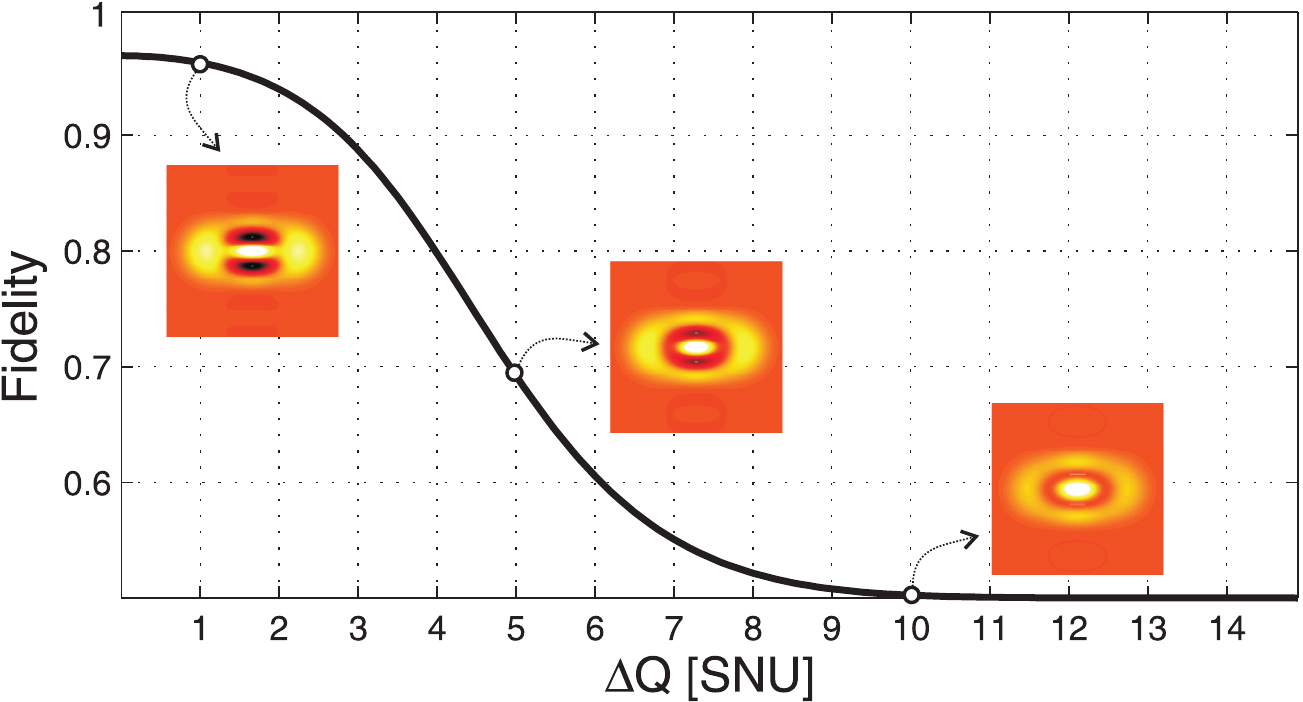}
  \caption{(Color online) Fidelity of the state amplified from two cats of size $\alpha = 1$ with an even cat of size $\beta = \sqrt{2}$ for a homodyning window $\Delta Q \in$ $]0, 15]$ SNU. The decrease in fidelity for wider homodyning windows is understandable from the increased overlap between the vacuum and the cat state, as illustrated in the inset of Fig. \ref{fig:amplification_setup}. The Wigner profile of the output is shown for three sample values of $\Delta Q$ at 1, 5, and 10 SNUs, respectively.}
\label{fig:HDwindowCatAmplificationFidelity}
\end{figure}

For another perspective on the dependence of fidelity on homodyning width, let's consider the amplification of two odd cat states of a fixed size $\alpha = 1$. The decrease in fidelity of the output with an even cat of size $\beta = \sqrt{2}$ is traced in Fig. \ref{fig:HDwindowCatAmplificationFidelity}. Also shown are three sample Wigner profiles of the output which display a clear degradation of the negativity as $\Delta Q$ increases.

The simulations presented above, as well as all other numerical results in this article are arrived at by a generic method of simulating linear transformation and projective measurements of states consisting of Gaussian superpositions (see Sec. \ref{sec:GaussianCircuitry}).

% Amplification of approximate small cats
%%%%%%%%%%%%%%%%%%%%%%%%%%%%%%%%%%%%%%%%%%%%%%%%%%%%%%%%%%%%%%%%%%%%%
\section{Amplification of approximate small cats}
\label{sec:KittenAmplification}

\begin{figure}[h]
  \includegraphics[width=.8\columnwidth]{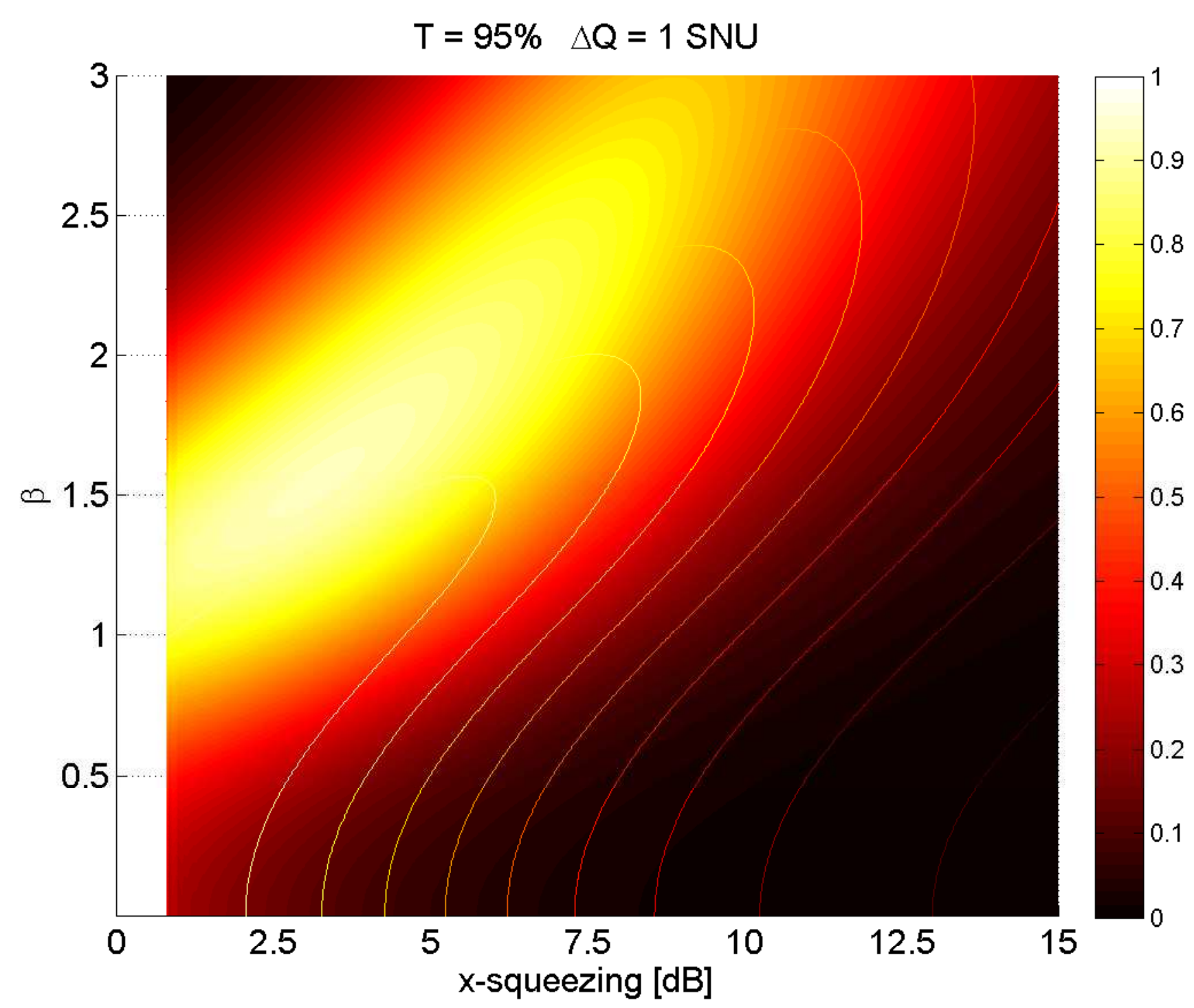}
  \caption{(Color online) Contour plot of the fidelity between an even cat $\ket{\kappa_{-}(\beta)}$ and the amplified PSSV obtained from an anti-squeezing $\xi_x$ between 0 and 15 dB. The transmission of the subtraction beam splitter is 95\%. The contour lines of PSSV generation fidelity are overlaid to better visualize the shift in target amplitude $\beta$ resulting from the amplification. The blank vertical stripe for $\xi_x \le 0.8$ dB is a region where numeric underflow is too frequent to produce reliable data. (This is because the normalization factor which enters in the fidelity is itself inversly proportional to the success probability. The latter tends to negligible values for small squeezing.)}
\label{fig:KittenAmplificationTrim_contour_DeltaQ_1_T_95}
\end{figure}

In this section, we consider the more realistic case where PSSVs are amplified, i.e., the inputs to Fig. \ref{fig:amplification_setup} are the outputs of Fig. \ref{fig:kitten_setup}. In Fig. \ref{fig:KittenAmplificationTrim_contour_DeltaQ_1_T_95}, the fidelity profile with an even cat of size $\beta$ is plotted with respect to the input (pure) squeezing for $\Delta Q = 1$ SNU and $T = 95\%$. The contour lines of PSSV generation are overlaid to visualize the amplification, i.e., the shift of the high-fidelity area upwards to larger values of $\beta$---cf. Fig. \ref{fig:KittenFidelity}.

\begin{figure}[h]
  \includegraphics[width=1\columnwidth]{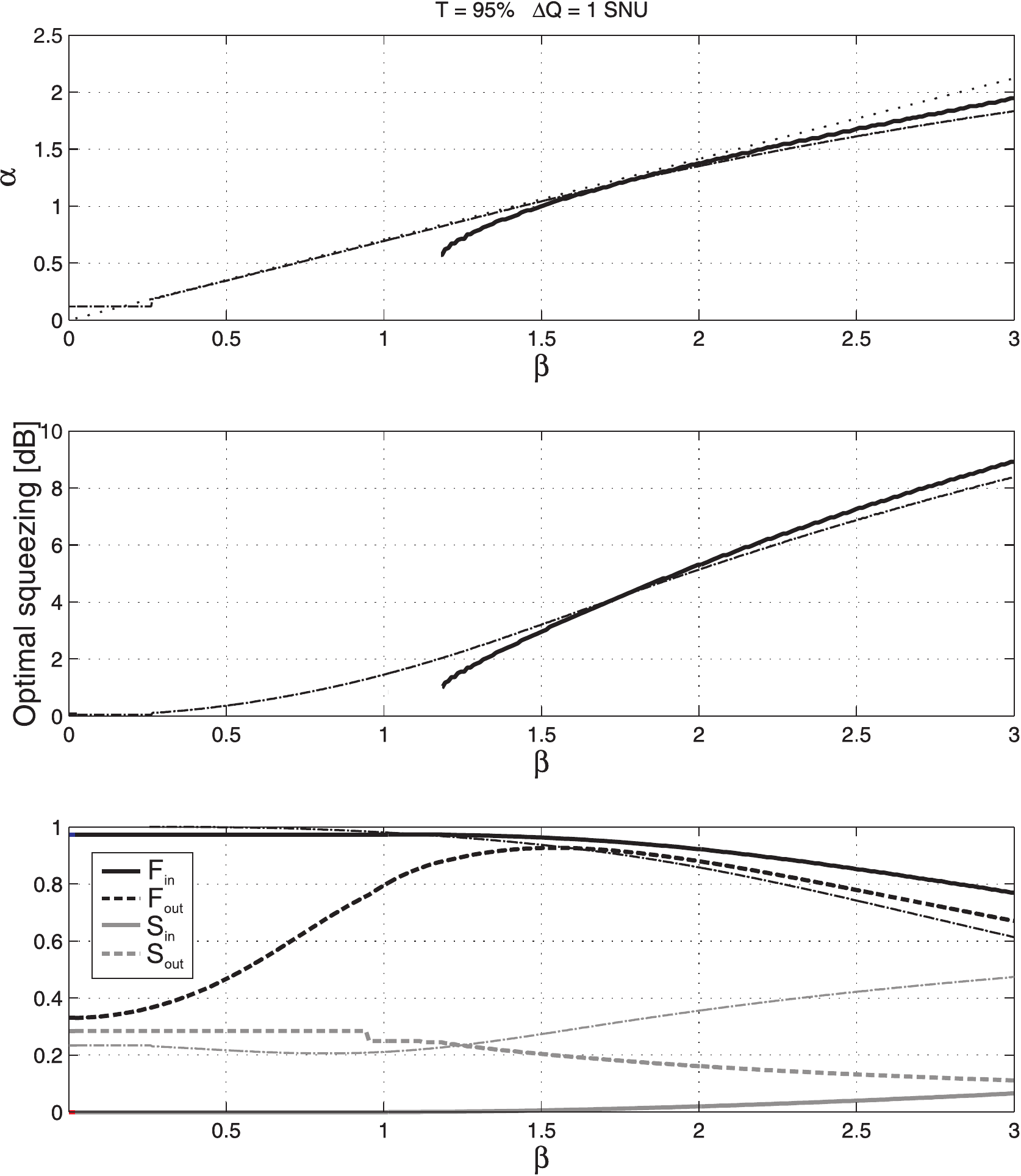}
  \caption{\textit{Top}: Dependence of the effective size $\alpha$ of the PSSV state input on the target size $\beta$ of the output such that the fidelity of the output with $\ket{\kappa_{+}(\beta)}$ is maximized. The dotted line marks the $\frac{\alpha}{\beta} = \frac{1}{\sqrt{2}}$ amplification ratio. The subtraction beam splitter for the PSSV generation is set to 95\% and the homodyning window to 1 SNU. \textit{Middle}: Pure squeezing required of the input in order to achieve the maximal fidelity with an even cat of size $\beta$. \textit{Bottom}: Maximum fidelity obtainable at the output with an even cat of size $\beta$ (dashed black), along with the corresponding success probability of the amplification (dashed gray). Also shown is the maximum fidelity of the required input PSSV with an odd cat  of size $\alpha$ (solid black), and the success probability of the PSSV generation (solid gray). (In all three graphs, the finer dash-dotted lines refer to the results of Lund \textit{et al.} in \cite{Lund2004}, with, in the bottom graph, the black and gray shadings representing fidelity and success probability, respectively.)}
\label{fig:BackwardOptimization}
\end{figure}

In order to assess the performance of the amplification scheme, we shall set a target even cat state of amplitude $\beta$ and retrace the quantum circuit to see what input squeezing is necessary to achieve the highest fidelity with $\ket{\kappa_{+}(\beta)}$ at the output. This is shown in Fig. \ref{fig:BackwardOptimization}, along with the dependence of the success probability and fidelity on $\beta$, as well as the size $\alpha$ of the odd cat that best matches the input PSSV. Let's assume, for instance, that we want to produce an even cat of size $\beta = 1.5$. The required squeezing for the PSSV will then be around 2.9 dB, corresponding to a fidelity of 96.4\% with an odd cat state of size $\alpha \approx 1.0$. The output, however, will have a fidelity of 92.6\% with $\ket{\kappa_{+}(\beta = 1.5)}$ and the success probability of the amplification will be 20\% (notwithstanding the success probability of 0.6\% required to produce the ``offline'' PSSV).

Note that the solid curves in the first two plots of Fig. \ref{fig:BackwardOptimization} only start at $\beta \approx 1.2$. This is because below that threshold, the optimal squeezing lies in a numerically unstable region marked the blank band in Fig. \ref{fig:KittenAmplificationTrim_contour_DeltaQ_1_T_95}.

A first-hand observation to be drawn from Fig. \ref{fig:BackwardOptimization} is the similarity it shows with Fig. \ref{fig:CatAmplificationOptimal} for small target cat sizes. Both scenarios with ideal cats and PSSV inputs start out with a fidelity around 30-35\% which then increases to over 90\% for targets cats of size $\beta \approx 1.5$. Beyond this point, however, the performance of PSSV amplification starts to degrade, whereas that of ideal cats can be pursued to indefinitely large target sizes while retaining unit fidelity. One conclusion is therefore that the approximation of cat states from PSSVs can be used to generate amplified states of sizes up to $\beta = 1.5$ with practically the same fidelity as if one used ideal odd cat states as inputs. On the other hand, both Figs. \ref{fig:CatAmplificationOptimal} and \ref{fig:BackwardOptimization} exhibit the plateau in optimal input cat size---or in the case of PSSVs, \textit{effective} cat size---which we discussed in the previous section. In that region, $\beta \in [0, 1]$, the ideal input cat states or PSSVs cease to have any dependence on the target size $\beta$ and discrete states, namely single photons, become the optimal input state.

In addition to simulating our own amplification scheme, we have overlaid as finer dash-dotted curves the results of Lund \textit{et al.}, which we shall refer to as the LJRK scheme \cite{Lund2004}. Instead of using homodyning, they mix the heralding arm with a coherent state on a balanced beam splitter such that both emerging modes contain at least one photon if the heralded mode is amplified. Contrary to the homodyne method, the LJRK projection is unambiguous as it does not yield any residual vacuum component, unlike in  (\ref{eq:AmplifiedCat}). This explains why the fidelity in the LJRK scheme remains quasi-ideal for low target sizes $\beta$. Beyond $\beta \approx 1.5$, however, both methods are comparable in terms of fidelity and amplification factor. The homodyne method offers nonetheless experimental advantages over the photon detection of LJRK in that it does not suffer as much of quantum inefficiency or electronic noise. In particular, the need for coincident detection of photons in LJRK suppresses the success probability by the \textit{square} of the quantum inefficiency. (The dash-dotted curve in Fig. \ref{fig:BackwardOptimization} assumes ideal quantum efficiency.)

\begin{figure}[h]
    \includegraphics[width=.9\columnwidth]{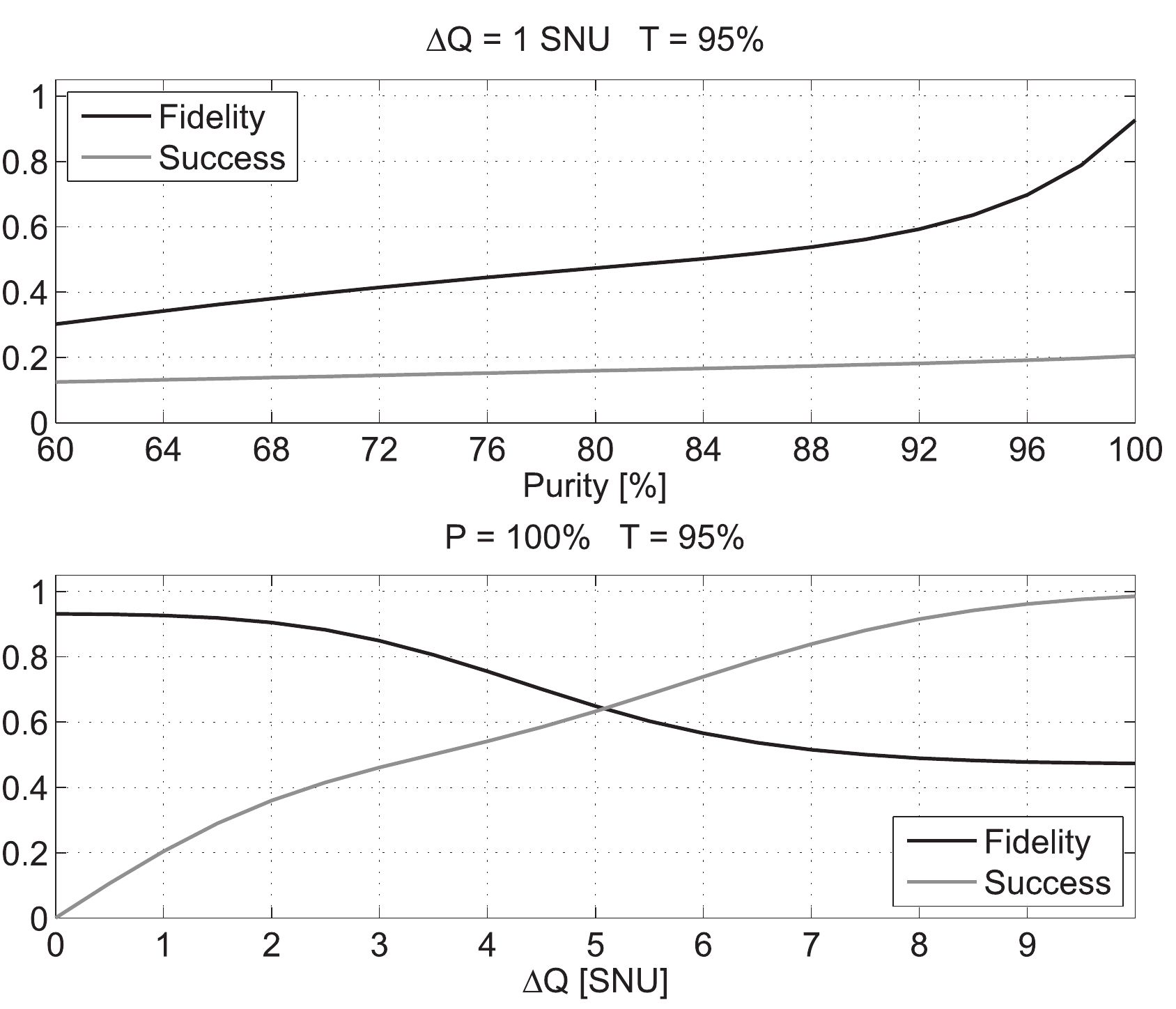}
  \caption{Robustness of the PSSV amplification to squeezing impurity (top) and homodyning width (bottom). The fidelity with an even cat of size $\beta = 1.5$ and the success probability of the amplification (assuming offline PSSVs) are plotted in black and gray, respectively. The transmission of the subtraction beam splitter from which the kittens are generated is 95\% and the anti-squeezing in \textit{p} is fixed $\xi_p = -2.9$. (This squeezing is chosen such that the fidelity with $\ket{\kappa_{+}(\beta = 1.5)}$ is maximized in the case of pure squeezing and $\Delta Q = 1$ SNU---cf. the middle plot of Fig. \ref{fig:BackwardOptimization}.)}
\label{fig:AmplifiedKittenRobustness}
\end{figure}

If one assumes ideal quantum efficiencies at the detectors, the two main nuisances to the performance of the amplification scheme are (i) the impurity $\mathcal{P}$ of squeezing, and (ii) the width $\Delta Q$ of the homodyning detection. To visualize the robustness of the scheme to these two factors, let's choose an optimal input squeezing with a target even cat of size, say, $\beta = 1.5$. From Fig. \ref{fig:BackwardOptimization}, this corresponds to $\xi_p = -2.9$ dB for a fidelity of 92.6\% and a success probability (assuming offline PSSVs) of 20\%. From this reference point, the trend of fidelity and success probability varying either $\mathcal{P}$ or $\Delta Q$ is plotted in Fig. \ref{fig:AmplifiedKittenRobustness}. Recall that, by convention, we model a decrease of purity by an increase in anti-squeezing ($\xi_x$) while maintaining squeezing proper ($\xi_p$) fixed.

% Scalability
%%%%%%%%%%%%%%%%%%%%%%%%%%%%%%%%%%%%%%%%%%%%%%%%%%%%%%%%%%%%%%%%%%%%%
\section{Scalability}
\label{sec:Scalability}

As was shown in the previous section, we are bound by a trade-off between amplification and fidelity. I.e., if the output fidelity is to be kept high, one has to work with smaller kittens to begin with, and thus cannot reach higher amplitudes (cf. Fig. \ref{fig:BackwardOptimization}). A high fidelity can however be achieved with larger states if we cascade the setups into a complete binary tree.  We shall discuss the scalability of such a recursive amplification in the case of perfect homodyning ($\Delta Q \rightarrow 0$) and pure squeezing ($\epsilon = 1$), both for ideal input cats and for PSSVs inputs.

Let's describe each iteration stage $k$ as the amplification of two identical wave functions $\varphi_{k}(x_1)$ and $\varphi_{k}(x_2)$ into a larger state $\varphi_{k+1}(x_1)$ where mode 2 has been post-selected by homodyning. This iteration step is given by
\beqa
& & \varphi_{k}(x_1) \cdot \varphi_{k}(x_2) \nonumber\\
& \rightarrow & \braces{\mbox{50:50 beam splitting}} \nonumber\\
& = & \varphi_{k}\tes{\frac{x_1 - x_2}{\sqrt{2}}} \cdot \varphi_{k}\tes{\frac{x_1 + x_2}{\sqrt{2}}} \nonumber\\
& \rightarrow & \braces{\mbox{projection on $\ketbra{x_2 = 0}{x_2 = 0}$}} \nonumber\\
& = & \varphi_{k}\tes{\frac{x_1}{\sqrt{2}}} \cdot \varphi_{k}\tes{\frac{x_1}{\sqrt{2}}} \nonumber\\
\varphi_{k+1}(x_1) & \propto & \varphi_{k}^2\tes{\frac{x_1}{\sqrt{2}}},
\eeqa
where the initial wave function $\varphi_{0}$ is that of the inputs. It can be obtained from the 1-variable analog of  ($\ref{eq:InnerProd}$). In the case of input PSSVs, this involves the highly unbalanced mixing two vacua, one of which is squeezed, and---ideally---a one-photon projector $\varphi_{\sub{$\ket{1}$}}$. I.e.,
\beqa
\varphi_{0}(x_1) = \lim_{T \to 1} \int_{-\infty}^{\infty} \! & & \varphi_{\sub{$\hat{S}\ket{0}$}}\tes{\sqrt{T}x_1 - \sqrt{1-T}x_2} \nonumber\\
& & \cdot \varphi_{\sub{$\ket{0}$}}\tes{\sqrt{1-T}x_1 + \sqrt{T}x_2} \nonumber\\ 
& & \cdot \varphi_{\sub{$\ket{1}$}}^{*}(x_2) \, \mathrm{d}x_2.
\label{eq:PSSVwavefunction}
\eeqa
Since we only want to investigate how scaling behaves, however, we shall simplify the PSSV by a squeezed single photon,
\beqa
\lim_{T \to 1} \varphi_{0}(x) & = & \bra{x} \hat{S} \ket{1} = \frac{\sqrt{2}}{\pi^{1/4} s^{3/2}} x e^{-x^2/2s^2}
\eeqa
where $s = 10^{\xi/20}$ is the factor by which the quadrature phase space is re-scaled as a consequence of the pure squeezing (i.e., $x \to s x$ and $p \to p/s$).

The wave function of the state at the $k^{\mbox{\scriptsize th}}$ iteration is given in closed-form by
\beq
\varphi_{k}(x) = \frac{1}{\mathcal{N}^{(k)}} \varphi_{0}^{2k} \tes{2^{-k/2} x},
\label{eq:ClosedFormIteration}
\eeq
where 
\beq
\mathcal{N}^{(k)} = \hak{\int_{-\infty}^{\infty} \! \varphi_{0}^{2k} \tes{2^{-k/2} x} \,  \mathrm{d}x}^{\frac{1}{2}}
\eeq
is the normalization factor.

\begin{figure}[h]
    \includegraphics[width=1\columnwidth]{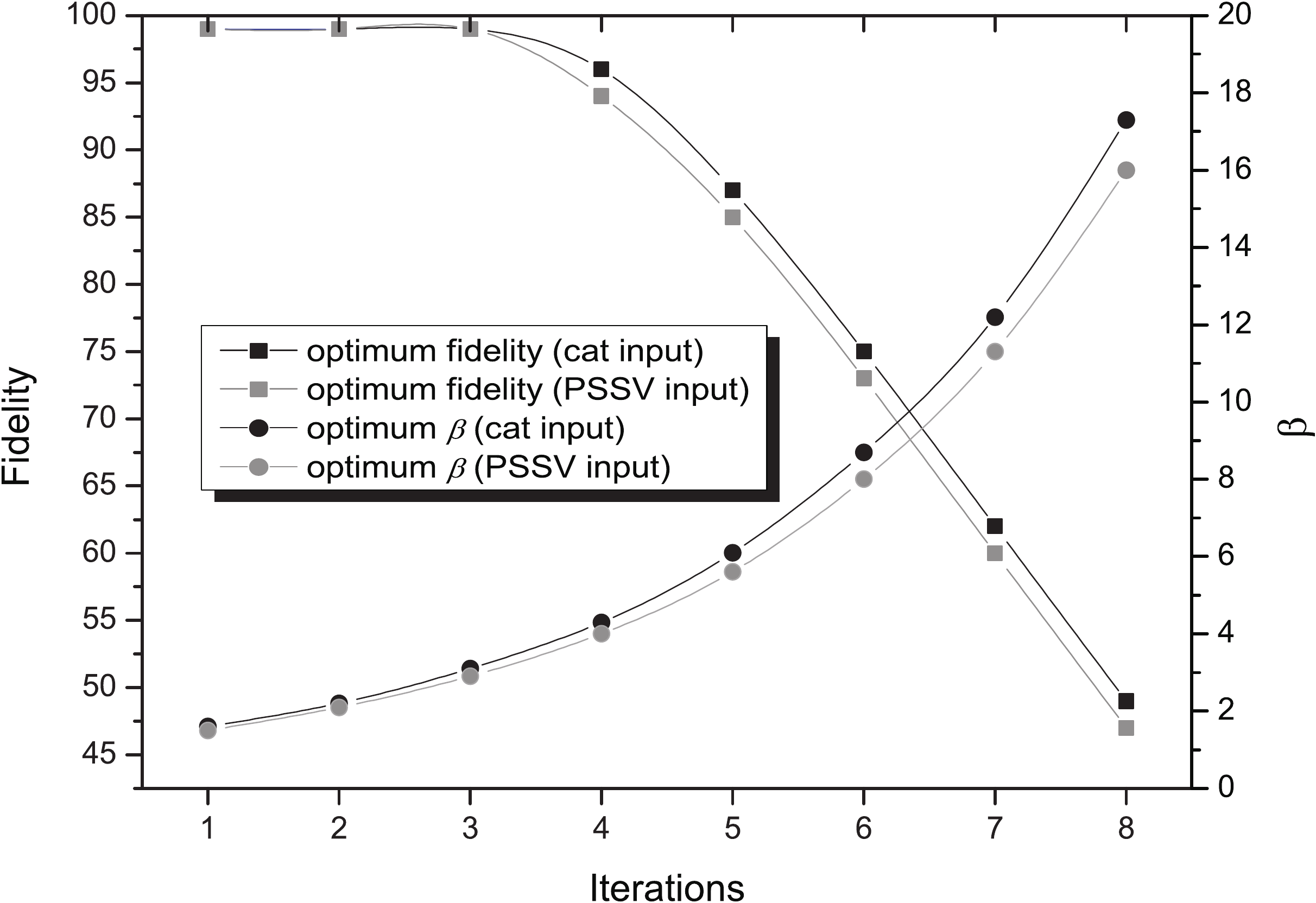}
  \caption{Maximum fidelity (squares) and corresponding effective size $\beta$  (disks) of the output as a function of the number of iterations. The inputs to the first iteration are an ideal odd cat of size $\alpha = 1$ (black) or a PSSV squeezed by 3 dB (gray).}
\label{fig:FidelityVsIterations}
\end{figure}

Fig. \ref{fig:FidelityVsIterations} tracks both the amplitude and the fidelity of the outputs as a function of the number of iterations $k$. As already expected from (\ref{eq:ClosedFormIteration}), the output amplitude grows as $\sqrt{2}^k$. This does not imply however that iterations could be carried out indefinitely. For the particular choice of initial conditions plotted, $\alpha = 1$ for the cat or 3 dB of squeezing for the PSSV, the fidelity drops below 90\% at the fifth iteration. This shortcoming of recursive amplifications is due to the non-unitarity of the amplification in the coherent state basis. Looking back at the idealized case of (\ref{eq:AmplifiedCat}), we see that the amplification is not a straightforward mapping of a cat state onto a larger one, but instead introduces an extra vacuum term. This  vacuum, which arises from the intrinsic ambiguity of the homodyne projection, is also amplified along with the cat. Even if one does start the first iteration with an ideal cat, any subsequent iteration $k$ will inherit this vacuum component, which in turn will contaminate the following step ${k+1}$ with additional terms orthogonal to an even cat.

\begin{figure}[h]
    \includegraphics[width=.85\columnwidth]{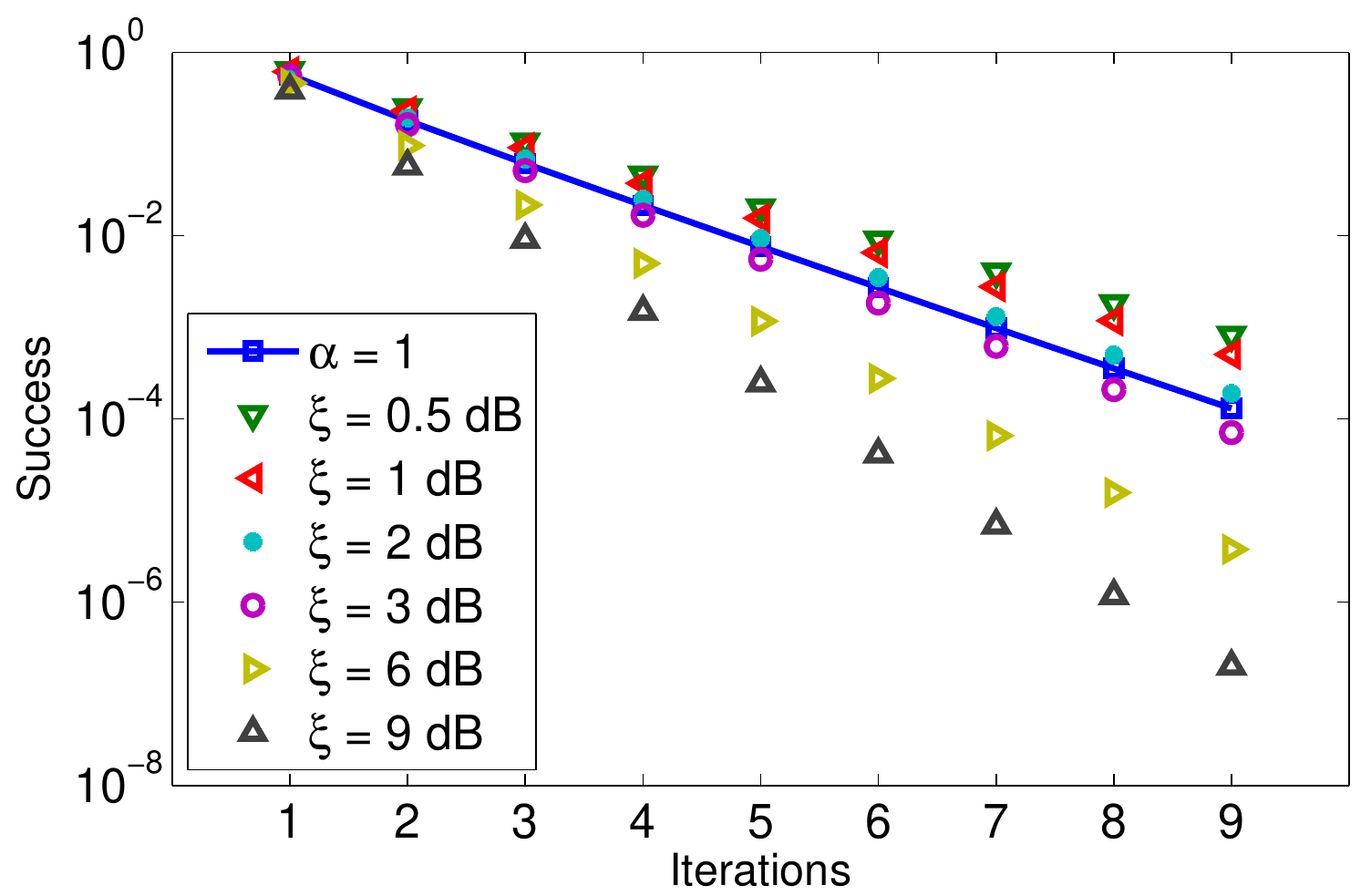}
  \caption{(Color online) Success probability of iterated amplifications for an input cat state $\ket{\kappa_{-}(\alpha = 1)}$ as well as for various input PSSVs with squeezing $\xi \in \braces{0.5, 1, 2, 3, 6, 9}$ dB.}
\label{fig:SuccessScaling}
\end{figure}

The success rate of iterated amplifications is the main issue facing scalability. As shown in Fig. \ref{fig:SuccessScaling}, the scheme exhibits a sharp drop in the success probability with the increasing number $k$ of iterations. Another obvious overhead is the number of input states which grows as $2^k$. 

% Conclusion
%%%%%%%%%%%%%%%%%%%%%%%%%%%%%%%%%%%%%%%%%%%%%%%%%%%%%%%%%%%%%%%%%%%%%
\section{Conclusion}
\label{sec:Conclusion}

We have presented an amplification protocol for cat states that is based on imprecise homodyne measurement. The performance of the scheme was assessed in terms of fidelity and success rate and illustrated by an optimized relation between the target size $\beta$ and the input size $\alpha$ of the cat states involved. Given that ideal cat states are challenging to produce, we also presented how the amplification  behaves with approximations to cat states, namely photon-subtracted squeezed vacuum. Here again, we determined the optimal relation between the input squeezing and the effective size of the output. The purity of squeezing at the input was determined to be crucial in achieving a high fidelity at the output. The amplification was however relatively robust to imprecise homodyne thresholding, thereby allowing an increase in success probability.

The recursive application of the amplification protocol is then simulated in the idealized case of exact homodyning and pure squeezing. Although the amplification factor does grow as as $\sqrt{2}^k$ with the number $k$ of recursions, the output fidelity eventually degrades due to the non-unitarity of the amplification. One is therefore constrained to a finite number of recursions where the state is amplified while at the same time retaining a high fidelity.

We saw in \S \ref{sec:KittenAmplification} that target cat sizes $\beta$ = 1.5 can be reached with fidelities up to 93\% if one uses PSSVs obtained from a 5\% tapped-off squeezed vacuum. (If one assumes ideally squeezed single photons, that fidelity can even increase to 98\%, cf. the first iteration in Fig. \ref{fig:FidelityVsIterations}. For higher iterations of the amplification, amplitudes of $\beta \approx 3$ can be obtained while maintaining  fidelity around 98\%.)  Although such approximate states may not permit fault-tolerant quantum computation, they nonetheless allow for proof-of-principle experiments that require effective cat sizes larger than those of basic PSSVs. For instance, our amplification protocol can find uses in teleportation \cite{Neergaard-Nielsen2012} or some demonstrations of quantum gates \cite{Grangier2011, Tipsmark2011}. The question of whether the amplified states can be harnessed for any particular use in quantum information will require a feasibility study of its own that pays particular attention to the tradeoff between target sizes and fidelities.

Let's conclude with a final note on the practical challenges to our protocol. In addition to the issues proper to PSSV preparation (and already discussed in \cite{Ourjoumtsev2006, Gerrits2010, Neergaard-Nielsen2006, Ourjoumtsev2007, Wakui2007}), the key challenge facing the amplification protocol is phase stability. This arises because the pairs of interfering PSSVs need to be synchronous, thereby leading to very small success rates. This is to be factored on top of the already small success rate of an otherwise ideal post-selection (Fig. \ref{fig:SuccessScaling}). Small success rates are compensated for by running the experiment over a larger batch of input states. However, this in turns requires that the relative phases---of the two interfering PSSVs and of the local oscillator---be kept stable for protracted periods using a particularly reliable locking system.

% Appendix
%%%%%%%%%%%%%%%%%%%%%%%%%%%%%%%%%%%%%%%%%%%%%%%%%%%%%%%%%%%%%%%%%%%%%

\appendix
\section{Mathematical derivations}

%  Expression for the purity of squeezed vacuum
\subsection{Expression for the purity of squeezed vacuum}
\label{sec:PuritySqueezedVacuum}

Let the variance of the vacuum phase space distribution be labeled by $V_0$. As will be discussed more formally in \S \ref{sec:GaussianCircuitry}, the squeezing operation along a quadrature $q$ consists of a re-scaling of phase space $q \rightarrow s_q q$ where $s_q = \sqrt{V_q/V_0}$ such that the new variance along $q$ is $V_q$. The relationship between the dimensionless re-scaling factor $s_q$ and the squeezing $\xi_q$ (in decibels) is given by
\beq
\xi_q = 10\log_{10}\frac{V_q}{V_0} = 20 \log_{10} s_q \Leftrightarrow s_q = 10^{\xi_q/20}. \label{eq:SqueezingVariables}
\eeq
The purity of a state of Wigner function $W$ is given by 
\beq
\mathcal{P} = 2\pi \iint \! W^2 \, \mathrm{d}x \mathrm{d}p, \label{eq:PurityDefinition}
\eeq
and the Wigner function of a squeezed vacuum state $\hat{S}\ket{0}$ is
\beq
W_{\sub{$\hat{S}\ket{0}$}}(x,p) = \frac{1}{\pi s_x s_p} e^{-(x/s_x)^2 - (p/s_p)^2}.
\eeq
The purity of squeezed vacuum can therefore be shown to be
\beq
\mathcal{P} = \frac{1}{s_x s_p} = 10^{-\frac{1}{20}\tes{\xi_x + \xi_p}}.
\eeq

% Error in the discrimination between the vacuum and a cat state with a homodyne projector
\subsection{Error in the discrimination between the vacuum and a cat state with a homodyne projector}
\label{sec:AmbiguousProj}

As mentioned in \S \ref{sec:CatAmplification}, the purpose of the homodyne measurement is to collapse the output state onto an even cat upon the detection of vacuum; cf. Eq. (\ref{eq:PsiOutAfterBS}) and the inset of Fig. \ref{fig:amplification_setup}. The projector for this measurement is
\beq
\Pihat_{\sub{\ket{0}}}^{\sub{hd}} = \int\limits_{-\frac{1}{2}\Delta Q}^{\frac{1}{2}\Delta Q} \! \ketbra{x}{x} \, \mathrm{d}x.
\eeq

\begin{figure}[h]
    \includegraphics[width=.9\columnwidth]{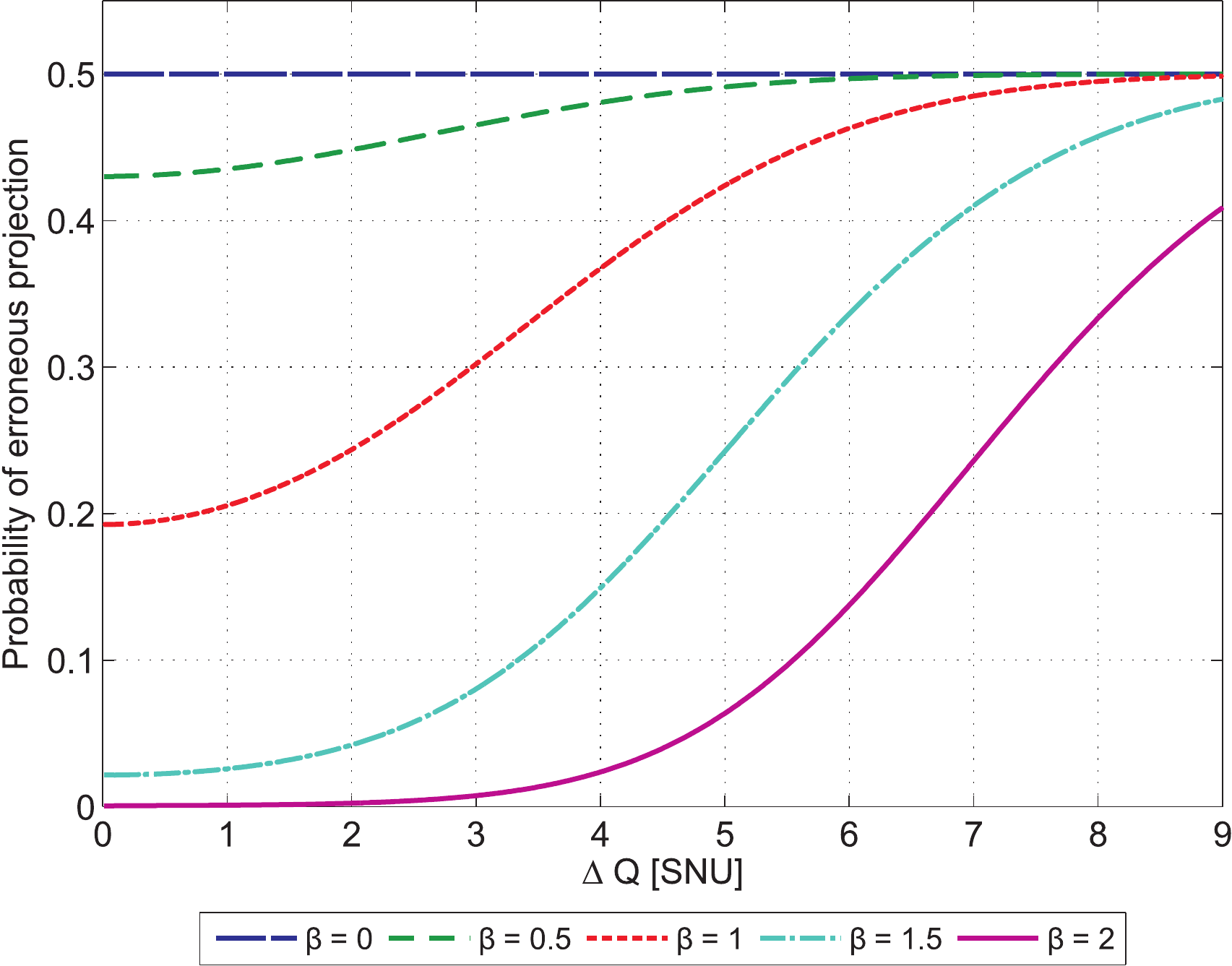}
  \caption{(Color online) Probability $P_{\sub{err}}$ of mistaking an even cat state of size $\beta$ for a vacuum state as a function of the quadrature acceptance window $\Delta Q$.}
\label{fig:Perr}
\end{figure}

Due to the intrinsic overlap of the wave functions of $\ket{0}$ and $\ket{\kappa_{+}(\alpha)}$, $\Pihat_{\sub{\ket{0}}}^{\sub{hd}}$ can only act as an  approximate discriminator between them. In addition, the finite width of the quadrature-selection window introduces an approximation of its own. The error in the discrimination, namely the probability of mistaking an even cat for the vacuum is given by
\beq
P_{\sub{err}} = \frac{P(\ket{\kappa_{+}(\beta)})}{P(\ket{\kappa_{+}(\beta)}) + P(\ket{0})},
\eeq
where, $P(\ket{\gamma}) = \mbox{Tr}\braces{\Pihat_{\sub{\ket{0}}}^{\sub{hd}} \ketbra{\gamma}{\gamma}}$ and $\ket{\gamma} \in \braces{\ket{0}, \ket{\kappa_{+}(\beta)}}$. Fig. \ref{fig:Perr} shows that the discrimination is best achieved for small $\Delta Q$ and large $\beta$. The asymptotic convergence to $P_{\sub{err}} = 0.5$ at large $\Delta Q$ or small $\beta$ indicates a complete lack of discrimination between the states: They become equally likely to be inferred by the homodyne projector.

%  Linear transformations and measurements with Gaussian states and operations
\subsection{Linear transformations and measurements with Gaussian states and operations}
\label{sec:GaussianCircuitry}

In this section, we shall present the calculational tools we have used to perform our simulations. All the quantum states and operations involved in this article are made up of Gaussian superpositions in the Wigner picture. I.e., any state $\rhohat$ or measurement operator $\Pihat$ can be written in the Wigner picture as a function 
\beq
W(x, p) = 
\sum_{i} G_i(x, p),
\label{eq:GenericGaussianWigner}
\eeq
where $G$ is a Gaussian function in both quadratures
\beq
G(x, p) = g_0 \, \exp{\tes{g_{1}x^2 + g_{2}x + g_{3}xp + g_{4}p^2 + g_{5}p + g_{6}}}
\label{eq:GenericGaussian}
\eeq
with constant coefficients $g_k \in \mathbb{C} : k \in \braces{0,\cdots,6}$ and $g_1, g_4 \in \mathbb{R}^-$. For example, a cat state is made up of four such Gaussians
\beqa
W_\sub{\ket{\kappa_{\pm}(\gamma)}}(x,p) & = & \sum_{j=1}^{4} G_{\kappa_{\pm}}^{(j)}(x,p,\gamma),
\eeqa
where
\begin{subequations}
\begin{align}
G_{\kappa_{\pm}}^{(1)}(x,p) &= -\frac{e^{-x^2 -p^2 +2\sqrt{2}i\gamma p}}{2\pi\tes{-e^{-2\gamma^2} \mp 1}}, \\
G_{\kappa_{\pm}}^{(2)}(x,p) &= -\frac{e^{-x^2 -p^2 -2\sqrt{2}i\gamma p}}{2\pi\tes{-e^{-2\gamma^2} \mp 1}}, \\
G_{\kappa_{\pm}}^{(3)}(x,p) &= \frac{e^{-x^2 -p^2 + 2\sqrt{2}\gamma x -2\gamma^2}}{2\pi\tes{-e^{-2\gamma^2} \mp 1}}, \\
G_{\kappa_{\pm}}^{(4)}(x,p) &= \frac{e^{-x^2 -p^2 - 2\sqrt{2}\gamma x -2\gamma^2}}{2\pi\tes{-e^{-2\gamma^2} \mp 1}},
\end{align}
\end{subequations}
whereas the APD operation $\Pihat^{(\mbox{\scriptsize APD})} = \hat{\mathbb{I}} - \ketbra{0}{0}$ is made up of only two Gaussians
\beq
W_{\Pihat^{(\mbox{\scriptsize APD})}}(x,p) = \sum_{j=1}^2 G_{\Pihat^{(\mbox{\scriptsize APD,\textit{j}})}}(x,p)
\label{eq:APDWigner}
\eeq
where
\begin{subequations}
\begin{align}
G_{\Pihat^{(\mbox{\scriptsize APD,1})}}(x,p) &= \frac{1}{2\pi}, \\
G_{\Pihat^{(\mbox{\scriptsize APD,2})}}(x,p) &= - \frac{1}{\pi} e^{-x^2 - p^2}.
\end{align}
\end{subequations}

If we now assume that all transformations underwent by the state $\rhohat$ prior to measurement are linear, as is the case in the setups presented in this article, then the generic Gaussian form (\ref{eq:GenericGaussian}) is conserved and only the quadrature vector 
\beq
\vec{q} = (x_1, p_1, \cdots, x_N, p_N)^T
\eeq
of the $N$ modes involved is transformed by a linear mapping
\beq
\vec{q} \rightarrow \textbf{A} \cdot \vec{q},
\label{eq:LinTrans}
\eeq
where the $2N \times 2N$ matrix \textbf{A} is determined by the linear assemblage of passive elements making up the circuit, e.g.,  beam splitters, phase shifters, squeezers, etc.

\begin{figure}[h]
  \includegraphics[width=.85\columnwidth]{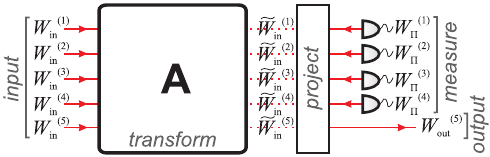}
  \caption{(Color online) Sketch of a black box quantum circuit made up of $N = 5$ input modes, $M = 4$ measurement modes, and $N - M = 1$ output modes. Each state occupying mode $k$ is represented by a Wigner function $W^{(k)}$. The input $\prod_n^N W_\sub{in}^{(n)}$ is mapped according to (\ref{eq:LinTransWigner}) into a linearly transformed state $\prod_n^N \tilde{W}_\sub{in}^{(n)}$ which is then projected onto a ``measurement state'' $\prod_m^M W_{\Pihat}^{(m)}$.} 
\label{fig:BlackBoxCircuit}
\end{figure}

Consider the generic quantum circuit depicted in Fig. \ref{fig:BlackBoxCircuit} where once again we only have Gaussian states and operators as defined by (\ref{eq:GenericGaussianWigner}-\ref{eq:GenericGaussian}). The effect of a re-scaled quadrature space (\ref{eq:LinTrans}) on a given Wigner function is equivalent to leaving the space unchanged while submitting the function to the inverse transformation. We shall label the transformed multimode Wigner function by a tilde such that the mapping (\ref{eq:LinTrans}) yields
\beq
W(\vec{q}) \rightarrow W(\textbf{A}^{-1} \vec{q}) = \tilde{W}(\vec{q}).
\label{eq:LinTransWigner}
\eeq

If we now apply the measurement operators over $M < N$ modes, the output over the remaining $N{-M}$ modes is given by 
\beq
W_{\rhohat_\sub{out}} = \iint\limits_{\mathbb{R}} \! \prod_{n}^{N} \tilde{W}_{\rhohat_\sub{in}}^{(n)} \cdot  \prod_{m}^{M} W_{\Pihat}^{(m)} \,  \mathrm{d}x_m \mathrm{d}p_m
\label{eq:InnerProd}
\eeq
where the superscript denotes the $k^{\mbox{\scriptsize th}}$ mode.

Note that the number of output modes equals the number of input modes $N$ minus the number of measured modes $M$. If $M = N$, then there is no output state and the inner product (\ref{eq:InnerProd}) leaves us with a scalar representing the success probability---or norm---of the projection of input states onto the measurement operators. Alternatively, this number could be interpreted as the fidelity between the overall states at either side of the diagram in Fig. \ref{fig:BlackBoxCircuit}.

The rationale for (\ref{eq:InnerProd}) is valid regardless of whether the Wigner functions are Gaussian superpositions: From a mathematical standpoint, projective measurements are inner products between the measured state and the measuring operator. In functional analysis, this translates to an integral of the product of two states---the measured and the measuring state---over the entire phase space of measurement. The simplicity that comes from using (\ref{eq:InnerProd}) with Gaussian superpositions is due to the fact that products and integrals of Gaussians are also still Gaussian. 

%  Homodyne measurement
\subsection{Wide homodyne measurement}
\label{sec:WideHomodyning}

A homodyne measurement corresponds to a projection on a given quadrature $\ketbra{q_\sub{0}}{q_\sub{0}}$ where $\ket{q_\sub{0}} = \cos{\theta} \ket{x_\sub{0}} + \sin{\theta} \ket{p_\sub{0}}$. This projector, when applied to a state whose Wigner function is $W(x,p)$, ``picks out'' the cross-section along $q_\sub{0}$ to yield a probability density. Such a projection is given in the quadrature basis by
\beq
\Pihat_\sub{HD}(q_\sub{0}) = \ketbra{q_\sub{0}}{q_\sub{0}},
\eeq
which, in the Wigner representation would most appropriately be given by the Dirac delta function
\beq
W_\sub{HD}(q_\sub{0}) = \delta(q-q_\sub{0}).
\label{eq:IdealHDWigner}
\eeq
If we choose for simplicity that $q_\sub{0} = x_\sub{0}$, then the projection operation (\ref{eq:InnerProd}) onto some state $W(x,p)$ yields
\beqa
& & \iint\limits_{\mathbb{R}} \! W(x,p) \cdot\ W_\sub{HD}(q_\sub{0}) \, \mathrm{d}x \, \mathrm{d}p \nonumber\\
& = & \iint\limits_{\mathbb{R}} \! W(x,p) \cdot \delta(x-x_\sub{0}) \, \mathrm{d}x \, \mathrm{d}p \nonumber\\
& = & W(x_\sub{0},p) \nonumber\\
& = & P(x_\sub{0}).
\eeqa

Intuitively, the probability of measuring exactly the quadrature $q_\sub{0}$ should be vanishingly small. Yet, $P(x_\sub{0})$ has a finite value since it is a probability \textit{density}. To model the more realistic scenario where homodyne detection has a given resolution bandwidth $\Delta q$ around the measured value $q_\sub{0}$, an interval projector is more appropriate
\beq
\Pihat_\sub{HD}(q_\sub{0},\Delta q) = \int_{q_\sub{0}-\frac{\Delta q}{2}}^{q_\sub{0}+\frac{\Delta q}{2}} \! \ketbra{q}{q} \, \mathrm{d}q
\eeq
or, in Wigner representation
\beqa
& & W_\sub{HD}(q_\sub{0},\Delta q) \nonumber\\
& = & \frac{1}{2\pi} \int_{q_\sub{0}-\frac{\Delta q}{2}}^{q_\sub{0}+\frac{\Delta q}{2}} \!  \delta(q-q') \, \mathrm{d}q' \nonumber\\
& = & \frac{1}{2\pi}\hak{\Theta\tes{q-q_\sub{0}+\frac{1}{2}\Delta q}-\Theta\tes{q-q_\sub{0}-\frac{1}{2}\Delta q}} \nonumber\\
\label{eq:WideHDWigner}
\eeqa
where $\Theta$ is the Heaviside step function. Note how (\ref{eq:WideHDWigner}) does not fit the Gaussian form of (\ref{eq:GenericGaussian}). For calculational ease, the wide homodyning operation is therefore best performed last so as to maintain the convenient Gaussianity of the quantum states as long as possible.

In the limiting case of ideal homodyning, $\Delta Q \rightarrow 0$, we expect (\ref{eq:IdealHDWigner}) and (\ref{eq:WideHDWigner}) to yield the same result.

% Bibliography
%%%%%%%%%%%%%%%%%%%%%%%%%%%%%%%%%%%%%%%%%%%%%%%%%%%%%%%%%%%%%%%%%%%%%

\end{document}